\documentclass[runningheads]{llncs}

\usepackage{graphicx}
\usepackage{tabularx}
\usepackage{threeparttable}
\newcommand{\mtnote}[1]{\textsuperscript{\TPTtagStyle{#1}}}
\usepackage{url}
\usepackage{enumitem}
\setlist[itemize]{nosep,leftmargin=0pt,topsep=4pt}

\begin{document}

\title{Procedure Model for Building Knowledge Graphs for Industry Applications}
\titlerunning{Procedure Model for Knowledge Graphs in Industry}
\author{Sascha Meckler}
\authorrunning{Meckler}
\institute{Fraunhofer IIS, Fraunhofer Institute for Integrated Circuits IIS, Germany\\
\email{sascha.meckler@iis.fraunhofer.de}}
\maketitle
\begin{abstract}
Enterprise knowledge graphs combine business data and organizational knowledge by means of a semantic network of concepts, properties, individuals and relationships. The graph-based integration of previously unconnected information with domain knowledge provides new insights and enables intelligent business applications. However, knowledge graph construction is a large investment which requires a joint effort of domain and technical experts. This paper presents a practical step-by-step procedure model for building an RDF knowledge graph that interconnects heterogeneous data and expert knowledge for an industry use case. The self-contained process adapts the "Cross Industry Standard Process for Data Mining" and uses competency questions throughout the entire development cycle. The procedure model starts with business and data understanding, describes tasks for ontology modeling and the graph setup, and ends with process steps for evaluation and deployment.
\keywords{knowledge graph construction (KGC) \and ontology engineering \and process model}
\end{abstract}

\section{Introduction}
The hype around knowledge graphs (KG) was started by the big US tech companies Google, Microsoft, Facebook, eBay and IBM who build large-scale graph-based solutions for their business models \cite{EKGbigtech}. (Enterprise) knowledge graphs are used for improved search, recommendations and personal/conversational agents, for targeted advertising, business analytics and risk assessment as well as for many other business use cases in e-commerce, social networks, finance, and other industries \cite{knowledgegraphs}.
In the last years, KG have been introduced to applications from manufacturing and Industry 4.0. Siemens uses an "Industrial Knowledge Graph" for digital twins of buildings or for monitoring machines and production processes in a factory \cite{SiemensKG}. In a study about the quality management of electronic products, Bosch implemented a KG solution that improved the efficiency of data analysis by 70\% \cite{BoschKG}.
Knowledge graphs are applied to scenarios that involve integrating, managing and extracting value from heterogeneous data sources, often called data silos. In applications where relations between entities are essential and flexibility is required, the graph-based abstraction of knowledge has advantages over conventional enterprise solutions which usually rely on a relational model \cite{knowledgegraphs}. General trends in the adoption of KG are the profitable use of graph structures for the integration of data from diverse sources at large scale, the combination of deductive formalisms (logical rules, ontologies, etc.) and inductive techniques (machine learning, analytics, etc.) to represent and accumulate knowledge \cite{knowledgegraphs}. The development of KGs usually follows a bottom-up strategy in which data is extracted, semantically enriched, integrated in a graph and continuously refined to improve the KG over time. 

Generic approaches for KG construction do not address the challenges and requirements of industry applications while domain-specific processes are hardly transferable to different application domains. At Fraunhofer IIS, we developed a procedure model for building knowledge graphs for industry applications (KG-PM). The agile procedure model is designed for the initial creation of an (enterprise) KG and its extension and evolution in subsequent iterations. The workflow was customized based on the experiences from research projects and consulting projects with industry partners. The model serves as a guideline for realizing individual KG projects with diverse application domains.
The KG-PM and its seven steps are presented in the following section. Section~\ref{section_relatedwork} describes existing process models and related work about the development of a KG.

\section{Procedure Model} \label{section_proceduremodel}
The procedure model for building knowledge graphs for industry applications (KG-PM) was derived from the Cross-Industry Standard Process for Data Mining (CRISP-DM) \cite{CRISP-DM}, a well-established reference process model for data mining projects. CRISP-DM is a methodology and a process model that organizes the life cycle of a data mining project into six phases -- business understanding, data understanding, data preparation, modeling, evaluation, and deployment. Since data mining and knowledge graph projects strive for data understanding and insights, the industry-, tool-, and application-neutral CRISP-DM model provided an ideal basis for the KG-PM.

We adapted CRISP-DM to the development of KG that build upon the standards of the Semantic Web, defined by the World Wide Web Consortium (W3C). The KG-PM is focused on the Resource Description Framework (RDF) \cite{w3c-rdf} as the data model for directed edge-labelled graphs. RDF Schema (RDFS) \cite{w3c-rdfs} and the Web Ontology Language (OWL) \cite{w3c-owl} are intended for modeling the data schema and semantics of the KG while using SPARQL\cite{w3c-sparql} as query language.
In comparison to labeled property graphs, RDF KG have advantages for data sharing and Web integration.

Within the workflow, we incorporated best practices such as the approved guide for ontology design by Noy and McGuinness \cite{ontology101}. Noy and McGuinness recommend the use of competency questions (CQ) as a way to determine the scope of an ontology and to evaluate the ontology concepts. Our model adopts and extends the usage of CQ, a list of use case-related questions which express the business goals and requirements from potential users of the KG. The CQ defined in the first step of the procedure model, support the entire KG building process from the start to the end. The questions define the scope of the KG, the ontology modeling and serve as a basis for the evaluation at the end of each development iteration.

\begin{figure}[h!]
    \centering
    \includegraphics[width=1.0\textwidth]{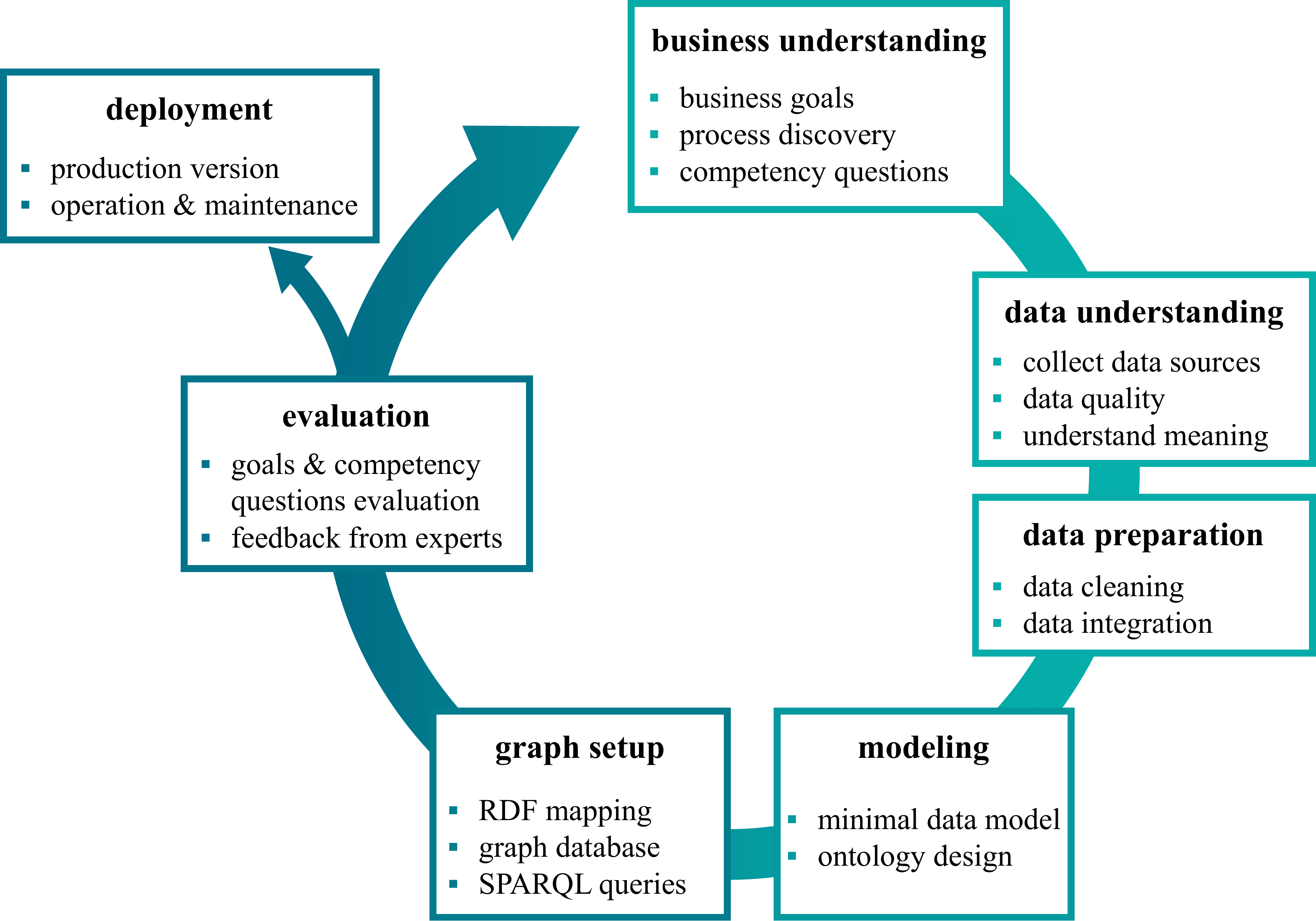}
    \caption{Procedure model for building an RDF KG for industry applications} 
    \label{fig_proceduremodel}
\end{figure}

\noindent Figure~\ref{fig_proceduremodel} visualizes the sequence and tasks of the steps in the process model. The KG-PM is tailored to suit requirements in industry applications, i.e. complex industry processes and diverse data silos with messy data. For this reason, the first step "business understanding" includes tasks for understanding the industry process, its stakeholders and relationships. The following steps "data understanding" and "data preparation" deal with low quality data from many heterogeneous, unconnected data sources. The workflow intentionally includes two steps for collecting and preparing data to draw attention to these tasks which are often underestimated. The data acquisition is designed for existing enterprise datasets with defined attributes. In contrast to the development of KGs from data on the Web, there is no need for open information extraction (OIE), the extraction of structured information from text documents using methods such as entity discovery, entity typing, entity linking or relation extraction. Once the data preparation is completed, the datasets are transformed into the RDF data model using the semantic concepts of the ontology which is developed in the "modeling" step. The ontology modeling for industry applications usually requires the design of a new ontology due to the special field of application. During the "graph setup", the generated RDF graphs are integrated in a graph database which supports RDF and the corresponding query language SPARQL. Based on the selected CQ from the first step, SPARQL queries are created, tested and evaluated in the phase "evaluation". The result of the evaluation step serves as input for the step "business understanding" of the next process iteration. If the evaluation of the KG instance proves its usability, steps are taken for deployment.
The following sub-sections describe the characteristics of each step with respect to the goals, work methods, involved stakeholders, the results and generated artifacts, and tool support. Additionally, difficulties and experiences from practical applications are explained.

\subsection{Business Understanding} \label{section_proceduremodel_business}
At the beginning of the KG construction, the industry use case and business goals are defined. In workshops with subject-matter experts, technical experts and potential users, brainstorming and creativity techniques are used to understand possible use cases, to define the scope and derive goals for the KG. In every subsequent development iteration, the use cases, scope and goals are updated. A summary of the characteristics of the step "business understanding" is given in table~\ref{table_step_business}.

\begin{table}[h!]
\begin{threeparttable}
\label{table_step_business}
\caption{Characteristics of the procedure step "business understanding"}
\centering
\setlength{\tabcolsep}{2pt}
\setlength{\parskip}{0pt}
\begin{tabular}{ m{0.23\textwidth} m{0.74\textwidth} }
\hline
Goals\newline &
\begin{itemize} 
    \item understand industry use case
    \item define business goals
    \item process discovery and process description (BPMN, EPC)
    \item find and prioritize competency questions for the goals
\end{itemize}
\\[-8pt]
\hline
Work methods\newline &
\begin{itemize} 
    \item group discussions
    \item creativity methods (e.g. 6-3-5 Brainwriting)
    \item (online) whiteboard brainstorming, clustering and ranking
\end{itemize}
\\[-8pt]
\hline
Stakeholders\newline &
\begin{itemize} 
    \item moderator
    \item domain experts
    \item technical experts, e.g. data stewards
\end{itemize}
\\[-8pt]
\hline
Results\newline &
\begin{itemize} 
    \item (clustered and ordered) list of competency questions
    \item (clustered and ordered) list of use cases
    \item process model diagram
\end{itemize}
\\[-8pt]
\hline
Tool support\newline &
\begin{itemize} 
    \item collaboration platforms such as Conceptboard\mtnote{1} or Miro\tnote{2}
    \item diagram editors (e.g. Diagrams.net\mtnote{3}, formerly Draw.io)
\end{itemize}
\\[-8pt]
\hline
\vspace{3pt}Difficulties\newline &
\begin{itemize} 
    \item prioritization of use cases and CQs
    \item differentiation between business questions and CQ
\end{itemize}
\\[-8pt]
\hline
\vspace{3pt}General advice\newline &
\begin{itemize} 
    \item prefer agile methods to classic requirements engineering
\end{itemize}
\\[-8pt]
\hline
\end{tabular}
\begin{tablenotes}
\vspace{5pt}
\scriptsize
\item[1] \url{https://conceptboard.com/}
\item[2] \url{https://miro.com/}
\item[3] \url{https://www.diagrams.net/}
\end{tablenotes}
\end{threeparttable}
\end{table}

\subsubsection{Process Discovery}
For understanding the industry use case, the KG-PM includes a task for process discovery. The business processes can be modeled with a simple flow chart like event-driven process chain (EPC) or using the Business Process Model and Notation (BPMN). If the connection of process steps to information systems is documented, the process model shows possible data sources for the KG. Furthermore, the interaction with domain experts or employees from production helps to understand the characteristics and requirements of the use case. During the process discovery task, business terms can be collected in a glossary or taxonomy which are useful inputs to the ontology design in section~\ref{section_proceduremodel_modeling}.

\subsubsection{Competency Questions}
We adopted the use of competency questions (CQ) from the ontology engineering process by Noy and McGuinness \cite{ontology101} who recommend CQ as a way to determine the scope of an ontology and to evaluate the ontology concepts later on. The use of CQs is extended to the entire KG constriction so that a list of CQ express the business goals and requirements from potential users of the KG. In workshops with an interdisciplinary team, the CQ are defined, clustered and prioritized. Complex business questions, which address multiple concepts of the ontology or the KG, are divided into smaller questions. At this point, it is often necessary to identify so-called "business questions" that cannot be answered by one or more queries to the KG because they would require statistical analysis or interaction with other information systems. If possible, the business questions are simplified into CQs or split into their query and analysis part. Furthermore, CQ should define clear criteria like threshold values to leave no room for interpretations.

\subsection{Data Understanding} \label{section_proceduremodel_dataunderstand}
The process step "data understanding" groups tasks for gathering data sources that are relevant for the selected use case and its CQs. Due to missing or mediocre data management or data governance, the identification and understanding of existing data sources requires more effort than expected. Use-case specific examination are often necessary to fully understand the content of poorly documented data sources. In an industry scenario, data sources not only include information from ERP systems, data warehouses or data lakes, but also data of production systems such as MES or SCADA. Table~\ref{table_step_dataunderstand} shows a summary of the characteristics of the step "data understanding".

\begin{table}[h!]
\begin{threeparttable}
\label{table_step_dataunderstand}
\caption{Characteristics of the procedure step "data understanding"}
\centering
\setlength{\tabcolsep}{2pt}
\setlength{\parskip}{0pt}
\begin{tabular}{ m{0.23\textwidth} m{0.74\textwidth} }
\hline
Goals\newline &
\begin{itemize} 
    \item collect available data sources
    \item check data quality
    \item understand meaning and relations (with domain experts)
    \item use case-specific examinations
\end{itemize}
\\[-8pt]
\hline
Work methods\newline &
\begin{itemize} 
    \item exploratory data analysis
    \item structured data quality evaluation (quality criteria such as unambiguous interpretability, uniform representation, credibility, faultlessness, completeness)
\end{itemize}
\\[-8pt]
\hline
\vspace{3pt}Stakeholders\newline &
\begin{itemize} 
    \item domain experts
    \item technical experts, e.g. data stewards
\end{itemize}
\\[-8pt]
\hline
\vspace{3pt}Results\newline &
\begin{itemize} 
    \item information about the relations between data (e.g. ERM or UML diagrams)
\end{itemize}
\\[-8pt]
\hline
Tool support\newline &
\begin{itemize} 
    \item data exploration tools such as OpenRefine\tnote{1}
    \item scripts for data analysis, e.g. using Jupyter\mtnote{2} notebooks in Python
\end{itemize}
\\[-8pt]
\hline
\vspace{3pt}Difficulties\newline &
\begin{itemize} 
    \item underestimated amount of manual work
    \item complex (relational) database schema without documentation or messy semi-structured data from spreadsheets
\end{itemize}
\\[-8pt]
\hline
\vspace{3pt}General advice\newline &
\begin{itemize} 
    \item analyze the quality of existing data sources to estimate the feasibility of the business goals early
\end{itemize}
\\[-8pt]
\hline
\end{tabular}
\begin{tablenotes}
\vspace{5pt}
\scriptsize
\item[1] \url{https://openrefine.org/}
\item[2] \url{https://jupyter.org/}
\end{tablenotes}
\end{threeparttable}
\end{table}

\subsubsection{Data Quality Evaluation}
The task of checking the data quality of the selected data sources is important to recognize problems at this early stage. In a structured data quality evaluation, each data source is assessed based on defined quality criteria. From 15 information quality criteria presented by \cite{information-quality-criteria}, we check data sources for unambiguous interpretability, uniform representation, credibility, faultlessness, and completeness.

\subsection{Data Preparation} \label{section_proceduremodel_dataprep}
The process step "data preparation" is strongly connected to the previous tasks for data understanding and data quality control. In this step, the data from the collected data sources is prepared for the mapping into the RDF graph data structure which takes place in the step "graph setup" (Sec.~\ref{section_proceduremodel_graphsetup}). Data preparation includes data cleaning, integration, restructuring and preprocessing operations. Details of this step of the procedure model are listed in table~\ref{table_step_dataprep}.

\begin{table}[h!]
\begin{threeparttable}
\label{table_step_dataprep}
\caption{Characteristics of the procedure step "data preparation"}
\centering
\setlength{\tabcolsep}{2pt}
\setlength{\parskip}{0pt}
\begin{tabular}{ m{0.23\textwidth} m{0.74\textwidth} }
\hline
\vspace{3pt}Goals\newline &
\begin{itemize} 
    \item process data from collected data sources
    \item prepare heterogeneous data for the mapping to RDF
\end{itemize}
\\[-8pt]
\hline
\vspace{3pt}Work methods\newline &
\begin{itemize} 
    \item data cleaning
    \item data integration (e.g. table joins)
\end{itemize}
\\[-8pt]
\hline
\vspace{3pt}Stakeholders\newline &
\begin{itemize} 
    \item technical experts, e.g. data engineers
\end{itemize}
\\[-8pt]
\hline
\vspace{3pt}Results\newline &
\begin{itemize} 
    \item CSV/TSV files or rel. database tables
\end{itemize}
\\[-8pt]
\hline
Tool support\newline &
\begin{itemize} 
    \item Python scripts (e.g. Pandas\mtnote{1} library and Jupyter\mtnote{2} notebooks)
    \item database clients, e.g. HeidiSQL\tnote{3}
\end{itemize}
\\[-8pt]
\hline
\vspace{3pt}Difficulties\newline &
\begin{itemize} 
    \item data preparation reveals data quality issues
    \item ETL data processing prior to the RDF mapping vs. declarative processing within RDF mapping rules
\end{itemize}
\\[-8pt]
\hline
\vspace{3pt}General advice\newline &
\begin{itemize} 
    \item "flat" or "denormalized" database tables can be mapped to RDF more easily
\end{itemize}
\\[-8pt]
\hline
\end{tabular}
\begin{tablenotes}
\vspace{5pt}
\scriptsize
\item[1] \url{https://pandas.pydata.org/}
\item[2] \url{https://jupyter.org/}
\item[3] \url{https://www.heidisql.com/}
\end{tablenotes}
\end{threeparttable}
\end{table}

\subsubsection{Denormalization}
Data preparation includes every operation for bringing the data exported from the selected data sources into a form that is well suited for mapping to RDF. Relational databases are usually structured in accordance with a series of so-called normal forms to reduce data redundancy and improve data integrity. Normalized relational databases are characterized by a great number of tables, which are connected via foreign key relations. This data structure is usually cumbersome for RDF mapping, because the attributes from different tables must be combined during the mapping. Although some RDF mapping tools allow for joining data based on key relations, it is easier to transform ”flat” or ”denormalized” tables, which contain every attribute of an entity in the same row/record.

\subsection{Modeling} \label{section_proceduremodel_modeling}
In the fourth step "modeling", the knowledge about the use case, the CQ and information from the previous data preparation steps are combined to create a minimal data model, for example an entity–relationship model of the important entities or concepts, their relationships and their properties. The activity of creating a preliminary model, the ontology conceptualization, is often done with the help of visual elements on paper or software. The minimal data model provides a basis for discussion with domain experts before starting the detailed ontology engineering process. The ontology has two purposes for the KG: On the data level, the RDFS or OWL ontology serves as a data schema for the creation of the RDF data graph, described in section~\ref{section_proceduremodel_graphsetup}. On a higher level, the ontology formalizes knowledge about the use case by defining axioms, relationships and constraints about domain concepts. A summary of the characteristics of the step "modeling" is given in table~\ref{table_step_modeling}.

\subsubsection{Ontology Development Methodology}
\begin{figure}[h]
    \setlength{\abovecaptionskip}{4pt}
    \setlength{\belowcaptionskip}{-12pt}
    \centering
    \includegraphics[width=1.0\textwidth]{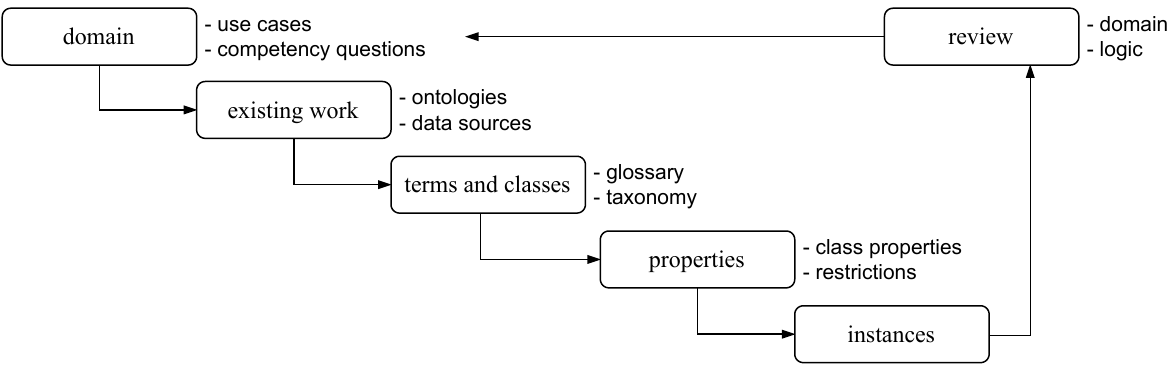}
    \caption{Steps for ontology development derived from \cite{ontology101}} 
    \label{fig_ontologysteps}
\end{figure}

The KG-PM refers to the well acknowledged guide for ontology development by Noy and McGuinness \cite{ontology101} who introduced an ontology design methodology for declarative frame-based systems. The steps in the iterative ontology-development process are illustrated in Fig.~\ref{fig_ontologysteps}. First, the domain and scope of the ontology is defined using competency questions. With the acquired domain knowledge, the development team should consider reusing existing ontologies. Important terms are then collected and organized into a class hierarchy. After that, the classes are complemented with properties and restrictions for the properties. In the last step, individual instances of classes are created. The state of the ontology is reviewed before starting a new iteration for extending and refactoring the domain ontology.

\begin{table}[h!]
\begin{threeparttable}
\label{table_step_modeling}
\caption{Characteristics of the procedure step "modeling"}
\centering
\setlength{\tabcolsep}{2pt}
\setlength{\parskip}{0pt}
\begin{tabular}{ m{0.23\textwidth} m{0.74\textwidth} }
\hline
\vspace{3pt}Goals\newline &
\begin{itemize}
    \item create a minimal data model
    \item design (domain) ontology
\end{itemize}
\\[-8pt]
\hline
Work methods\newline &
\begin{itemize} 
    \item domain modeling (bottom-up or top-down; classes and relations)
    \item ontology development
    \begin{itemize}[nosep,leftmargin=10pt,label={--}]
     \item following acknowledged steps by \cite{ontology101}
     \item use existing ontologies as a foundation or include them for specific parts of the ontology
     \end{itemize}
\end{itemize}
\\[-8pt]
\hline
Stakeholders\newline &
\begin{itemize} 
    \item domain experts
    \item expert for ontology modeling ("ontologist")
    \item (optional) technical/data experts
\end{itemize}
\\[-8pt]
\hline
\vspace{3pt}Results\newline &
\begin{itemize} 
    \item ontology visualization
    \item ontology RDF serialization
\end{itemize}
\\[-8pt]
\hline
Tool support\newline &
\begin{itemize} 
    \item visual modeling tools like Diagrams.net\mtnote{1} or yEd\mtnote{2} together with graphical notations, e.g. Graffoo~\cite{graffoo} or Chowlk~\cite{chowlk}
    \item text editor with RDF support, e.g. Visual Studio Code\mtnote{3} with the Stardog RDF Grammars extension\tnote{4}
    \item ontology modeling tool, usually Protégé\tnote{5}
    \item ontology documentation tools, e.g. Ontospy~\cite{ontospy} or WIDOCO~\cite{WIDOCO}
\end{itemize}
\\[-8pt]
\hline
\vspace{3pt}Difficulties\newline &
\begin{itemize} 
    \item correct integration of existing ontologies (e.g. no unintended entailments)
\end{itemize}
\\[-8pt]
\hline
\vspace{3pt}General advice\newline &
\begin{itemize} 
    \item use visual drafts for feedback before writing RDF serialization
    \item first priority of the ontology is the definition of a data schema for the RDF graph
\end{itemize}
\\[-8pt]
\hline
\end{tabular}
\begin{tablenotes}
\vspace{5pt}
\scriptsize
\item[1] \url{https://app.diagrams.net/}
\item[2] \url{https://www.yworks.com/products/yed}
\item[3] \url{https://code.visualstudio.com/}
\item[4] \url{https://github.com/stardog-union/stardog-vsc}
\item[5] \url{https://protege.stanford.edu/} 
\end{tablenotes}
\end{threeparttable}
\end{table}

\vspace*{-1.0\baselineskip}

\noindent Alternatively, the Linked Open Terms (LOT) methodology is a new methodology for building ontologies for industry projects \cite{LOT-ontology-eng}. The LOT methodology is based on existing methodologies, aligned with agile development techniques and Linked Data principles. The iterative workflow of LOT defines four activities and their artefacts: The result of the first activity for requirements specification (1) is the "ontology requirements specification document". After the ontology implementation (2) using a formal language like OWL and RDFS, the ontology artefact is published (3) online. The final ontology maintenance (4) includes activities for updating the ontology during its life cycle. Moreover, the LOT methodology provides suggestions for software support for the sub-activities of each step.

\subsubsection{Ontology Design Patterns}
Ontology design is a creative process that has many solutions \cite{ontology101}. There are no universal best practices that apply to every application domain. Dave McComb identifies eight different “schools of ontology design” with different characteristics and consensus: philosophy school, vocabulary and taxonomy school, relational school, object-oriented school, standards school, Linked Data school, NLP/LLM school and a data-centric school \cite{ontology-schools}. The team of ontologists should create a hybrid of one or more of these schools that provides the best solution to the objectives of the project \cite{ontology-schools}.
Despite that, the ontology engineering community has some agreements on bad practices or modeling errors. OOPS!~\cite{ontology-pifalls} is an online ontology pitfall scanner that is backed by a catalogue of typical modeling errors.
Methods for ontology evaluation include the comparison to a reference ontology known as the gold standard (1), the comparison to a text corpus that covers a given domain significantly (2), the evaluation based on certain criteria, e.g. for the structure or quality of an ontology (3), and the measurement how far an ontology helps improving the results of a certain task (4) \cite{Survey_OntoEval}. The KG-PM follows a task-based evaluation strategy in which the tasks are to answer the CQs. 

\subsubsection{Practical Ontology Modeling}
The ontology for an RDF KG is modeled using the ontology languages RDFS~\cite{w3c-rdfs} or OWL~2~\cite{w3c-owl}. OWL 2 has multiple subsets, so-called profiles, that trade off different aspects of OWL's expressive power in return for better computational or implementational benefits \cite{w3c-owl}. The choice of the OWL 2 profile depends on the requirements for complex knowledge modeling and reasoning versus query and reasoning performance. If ontology-based data integration is the main goal of the KG, a lightweight OWL 2 profile, which only provides limited expressions, should be used. In this case, the first priority of the ontology is the definition of a data schema for the RDF graph. The ontology can be limited to simple class hierarchies and relationships, for example using the subclass, domain and range properties from RDFS and ObjectProperty and DatatypeProperty from OWL.
Another practical way to avoid ontologies with complex OWL constructs is the complementary use of RDF shapes languages such as SHACL~\cite{w3c-shacl}. In contrast to the logical OWL restrictions, SHACL defines RDF shape constraints that are validated on a schema level.

In our experience, existing ontologies are typically not suitable for industry use cases with specialized requirements. However, there are several strategies for re-using existing work: General metadata attributes can be described with existing vocabularies or ontologies such as the Dublin Core Metadata Elements \cite{dublincore}. SKOS \cite{w3c-skos} can be used to include existing taxonomies, classification schemes, or thesauri for business terms. Furthermore, the ontology can re-use available ontologies for specific parts of the ontology, e.g. the W3C Organization Ontology \cite{w3c-org} for modeling organizational structures. A shared vocabulary like schema.org \cite{schemaorg} can be used as a foundation, from which more specialized classes or properties are derived. The complete ontology for the KG might contain new domain-specific concepts, reference industry taxonomies, import multiple existing ontologies and include alignments between these ontologies \cite{bosch-datalake-kg}.
The reuse of well-known ontologies has advantages even though enterprise KG are not destined for Linked Open Data. It not only reduces the effort for modeling, it simplifies understanding and sharing the KG with colleagues, partners or other branches of the company.

\subsubsection{Ontology Documentation}
The visualization of an ontology in a graph with nodes and edges has proven an efficient way for the discussion with domain experts and non-technical stakeholders. The graphical ontology notation Graffoo \cite{graffoo} defines elements for the visual modeling of OWL ontologies in free diagram editors like yEd or Diagrams.net. Alternatively, the Chowlk \cite{chowlk} framework can be used to generate an OWL RDF representation of a visual ontology conceptualization designed in Diagrams.net. The Chowlk visual notation covers high-level as well as fine-grained constructs from the OWL and RDFS language. The generated OWL ontology can be imported into ontology editing software like Protégé or used for generating documentation. Tools such as Ontospy~\cite{ontospy} or WIDOCO~\cite{WIDOCO} create textual and visual representations of an OWL/RDFS ontology which can be used for presenting and discussion the ontology with non-technical experts.

\subsection{Graph Setup} \label{section_proceduremodel_graphsetup}
The actual KG instance is built in the step "graph setup" which is characterized in table~\ref{table_step_graphsetup}. The data prepared for the KG (sect.~\ref{section_proceduremodel_dataprep}) is transformed into the RDF data format and thereby semantically enriched. Semi-structured or structured data is mapped to instances of classes with properties from the ontology developed in the step "modeling" (sect.~\ref{section_proceduremodel_modeling}). The generated RDF data graphs are then integrated in a RDF/Triple Store, a graph database that supports RDF and SPARQL. By using the W3C Semantic Web standards, the generated RDF is compatible with every RDF/Triple Store regardless of the mapping tool. Finally, SPARQL queries are used to check the integrity of the KG instance. In the second, third and following iterations of the procedure model, the graph setup is updated and extended.

\begin{table}[h!]
\begin{threeparttable}
\label{table_step_graphsetup}
\caption{Characteristics of the procedure step "graph setup"}
\centering
\setlength{\tabcolsep}{2pt}
\setlength{\parskip}{0pt}
\begin{tabular}{ m{0.23\textwidth} m{0.74\textwidth} }
\hline
Goals\newline &
\begin{itemize}
    \item create and execute RDF mapping rules
    \item integrate RDF data in a graph database
    \item test KG instance with queries
\end{itemize}
\\[-8pt]
\hline
Work methods\newline &
\begin{itemize} 
    \item select the right mapping tool (based on data types, interfaces, personal experience)
    \item write mapping rules and transform data into RDF
    \item install the graph database, create datasets and import RDF files
    \item validate the RDF data using SPARQL queries
\end{itemize}
\\[-8pt]
\hline
\vspace{3pt}Stakeholders\newline &
\begin{itemize} 
    \item technical experts for data and graph database
    \item IT experts for the IT infrastructure
\end{itemize}
\\[-8pt]
\hline
\vspace{3pt}Results\newline &
\begin{itemize} 
    \item RDF mapping rules
    \item RDF data graph
\end{itemize}
\\[-8pt]
\hline
Tool support\newline &
\begin{itemize} 
    \item RDF mapping tools, e.g. RMLmapper\mtnote{1} or Tarql\tnote{2}
    \item graph databases (RDF/Triple Stores), e.g. Apache Jena Fuseki\mtnote{3} or GraphDB\tnote{4}
\end{itemize}
\\[-8pt]
\hline
\vspace{3pt}Difficulties\newline &
\begin{itemize} 
    \item complicated RDF mapping rules if the ontology is very different from the current data(base) schema or if the mapping depends on conditions
\end{itemize}
\\[-8pt]
\hline
General advice\newline &
\begin{itemize} 
    \item mapping rules can be simplified with more preprocessing in the step "data preparation"
    \item CSV files are easily mapped with Tarql, but RML/RMLmapper supports interfaces other than files such as JDBC or Web APIs
\end{itemize}
\\[-8pt]
\hline
\end{tabular}
\begin{tablenotes}
\vspace{5pt}
\scriptsize
\item[1] \url{https://github.com/RMLio/rmlmapper-java}
\item[2] \url{https://github.com/tarql/tarql}
\item[3] \url{https://jena.apache.org/documentation/fuseki2/}
\item[4] \url{https://www.ontotext.com/products/graphdb/}
\end{tablenotes}
\end{threeparttable}
\end{table}

\vspace*{-1.0\baselineskip}

\subsubsection{RDF Mapping Tools}
The choice of the RDF mapping tool depends primarily on the type of data sources. There is a great variety of RDF mapping tools that differ in terms of
\begin{itemize}[leftmargin=24pt,topsep=2pt]
  \item the iterface (GUI or CLI),
  \item the definition of mapping rules (e.g. domain-specific language, SPARQL-based, annotations),
  \item ease of use and assistance to mapping rule generation,
  \item supported input data formats (e.g. tabular, streams, data cubes),
  \item ETL capabilities and integration with other systems (e.g. databases or Web APIs) and
  \item the license (e.g. open-source or part of a commercial product).
\end{itemize}
Van Assche et al. \cite{SurveyDeclarativeRDFGen} give an overview over mapping languages and materialization and virtualization systems for generating RDF graphs from heterogeneous (semi-)structured data in a declarative way.
The study categorizes and compares eight mapping languages: Dedicated mapping languages such as RML, xR2RML or D2RML extend the W3C-recommended RDB to RDF Mapping Language (R2RML) or use a custom syntax like D-REPR. Repurposed languages repurpose existing specifications for the definition of mapping rules. Query-language-driven mapping languages such as SPARQL-Generate, Facade-X, or XSPARQL repurpose the SPARQL query language  while constraint-driven mapping languages like the Shape Expressions Mapping Language (ShExML) built upon constraint languages for expressing mapping rules. The study of 19 materialization-based and 11 virtualization-based mapping systems showed that most systems work with their own introduced mapping language. However, eight systems were designed for RML and four systems work with a modified versions of R2RML\cite{SurveyDeclarativeRDFGen}.
The declarative definition of schema transformations are often combined with rules for data transformation which often depend on the used mapping language.
A study of over 20 systems supporting the conversion of tabular data to RDF has been done by Fiorelli and Stellato \cite{rdf-mapping-tabular}. In their comparison of approaches to the conversion, the types of inputs and data formats, easiness of use and integration features, Fiorelli and Stellato found that the majority of tools specialize on a row-based layout of the tabular data. The preprocessing of data from the selected data sources into a layout where each row represents one entity is part of the step "data preparation". 
We experienced good results with the use of Tarql for CSV/TSV files and the application of RML mappers for relational databases.

\subsubsection{RDF/Triple Stores}
The choice of the data management system for the generated RDF data depends on the technical requirements such as the estimated data volume, querying performance, or integration with other systems, but also on requirements for licenses and support.
Ali et al.~\cite{RDFStoresSPARQLengines} present a comprehensive survey of state-of-the-art storage, indexing and query processing techniques for evaluating SPARQL queries in a local setting and describe graph partitioning techniques for a distributed setting.
In data warehousing approaches, the entire RDF data is stored in one centralized RDF database which is usually deployed on a single machine. 
Approaches for distributed RDF processing include cloud-based solutions, partitioning-based approaches and federated SPARQL evaluation systems \cite{RDFStoresSPARQLengines}.
The extended version \cite{RDFStoresSPARQLengines-ext} of \cite{RDFStoresSPARQLengines} provides an overview of 135 local and distributed RDF stores and SPARQL query engines in terms of storage, indexing, join and query processing, and partitioning techniques. Moreover, references to SPARQL benchmarks and benchmarks for RDF stores are given \cite{RDFStoresSPARQLengines-ext}.
For large-scale semantic applications in industry scenarios, Modoni et al.~\cite{RDFStoreSurvey1} examined different RDF stores like AllegroGraph, OpenLink Virtuoso or Stardog with respect to security, versioning, and capabilities for streaming data and binary data. They conclude that the majority of the surveyed RDF stores do not yet support versioning and handling of streaming data in an effective way \cite{RDFStoreSurvey1}. Lam et al. recently presented a performance evaluation of five RDF stores on the basis of the complete Wikidata KG in which RDFox followed by GraphDB and Stardog had the best overall query performances \cite{WikidataStoreQEBenchmark}. In academic studies, Apache Jena TDB/Fuseki and OpenLink Virtuoso are often used for RDF storage \cite{KGDevPubSurvey}.
If the selected RDF/Triple Store has support for SHACL~\cite{w3c-shacl}, the RDF store is able to validate the shape of imported data automatically based on deposited SHACL rules.

\subsection{Evaluation} \label{section_proceduremodel_evaluation}
After constructing the first KG instance or updating an existing KG in the previous step, the KG is evaluated in terms of business goals and quality. A summary of the characteristics of the step "evaluation" is given in table~\ref{table_step_evaluation}.

\begin{table}[h!]
\begin{threeparttable}
\label{table_step_evaluation}
\caption{Characteristics of the procedure step "evaluation"}
\centering
\setlength{\tabcolsep}{2pt}
\setlength{\parskip}{0pt}
\begin{tabular}{ m{0.23\textwidth} m{0.74\textwidth} }
\hline
Goals\newline &
\begin{itemize}
    \item evaluation of goals and competency questions
    \item feedback from stakeholders
    \item KG quality evaluation
\end{itemize}
\\[-8pt]
\hline
Work methods\newline &
\begin{itemize} 
    \item iterate through the CQ from the step "business understanding"
    \begin{itemize}[nosep,leftmargin=10pt,label={--}]
         \item collect or create sub-questions
         \item try to create SPARQL queries for each sub-question
         \item rate the feasibility of the query and document requirements for later fulfillment
    \end{itemize}
    \item discuss the results of the CQ evaluation with stakeholders
    \item systematic KG quality control
\end{itemize}
\\[-8pt]
\hline
Stakeholders\newline &
\begin{itemize} 
    \item moderator
    \item domain experts
    \item technical experts for the KG
\end{itemize}
\\[-8pt]
\hline
\vspace{3pt}Results\newline &
\begin{itemize} 
    \item CQ evaluation table
    \item SPARQL queries for CQs
\end{itemize}
\\[-8pt]
\hline
\vspace{3pt}Tool support\newline &
\begin{itemize} 
    \item SPARQL client: visual query editor (e.g. YASGUI\mtnote{1}) or query builder
    \item tool for documenting the CQ evaluation, e.g. Microsoft Excel
\end{itemize}
\\[-8pt]
\hline
Difficulties\newline &
\begin{itemize} 
    \item CQs often include several sub-questions that require more than one SPARQL query
    \item missing details, e.g. threshold values, for creating queries from textual CQ
    \item business questions require statistical analysis
\end{itemize}
\\[-8pt]
\hline
General advice\newline &
\begin{itemize} 
    \item The answering of CQs can be demonstrated by means of a visualization of the ontology
\end{itemize}
\\[-8pt]
\hline
\end{tabular}
\begin{tablenotes}
\vspace{5pt}
\scriptsize
\item[1] \url{https://github.com/TriplyDB/Yasgui}
\end{tablenotes}
\end{threeparttable}
\end{table}

\noindent The CQs that have been selected in the step "business understanding", define the goals and user requirements for the current development phase. Thus, the achievement of the objectives is assessed on the basis of the answerability of the CQs by means of queries to the KG.
Each selected CQ is broken down into sub-questions that can be translated to SPARQL queries. The collected SPARQL queries are executed against the KG and their query results are examined and documented in an CQ evaluation table. The table might have columns for the CQ, its sub-questions, the corresponding SPARQL query, the rating of the query result, required work steps or notes. At the end of this systematic testing, this evaluation table gives an overview over achieved and open goals for the KG construction so that a fulfillment rate can be calculated.

\subsubsection{Evaluation Control Loop}
\begin{figure}[h]
    \centering
    \includegraphics[width=.75\textwidth]{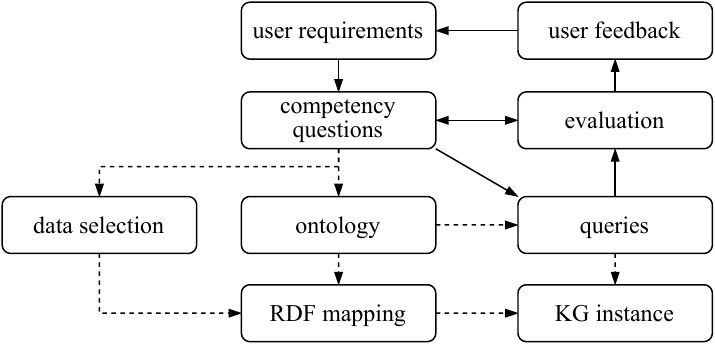}
    \caption{Control loop that interconnects user requirements, competency questions, queries and the evaluation process.} 
    \label{fig_controlloop}
\end{figure}

The business goals and the results of the systematic CQ evaluation are discussed with domain experts and other stakeholders. The remaining CQs can be categorized using a cost-benefit matrix that shows the relation between the technical work for implementing a specific CQ and the benefits for the users of the KG. After creating SPARQL queries for some of the CQs, technical experts can estimate the costs for extending the KG so that a new CQ can be answered while potential users reflect on the benefit based on the results of the current state of the KG.
The feedback from the evaluation is an input for the step "business understanding" of the following KG development iteration in which the goals and requirements are updated. This creates the control loop displayed in Fig.~\ref{fig_controlloop} where user requirements and goals are expressed as CQs and implemented in the KG. The user feedback from the evaluation of the queries derived from the CQs is used to update the initial goals and requirements.
This strategy can be extended to a "test-driven knowledge graph construction" \cite{TD-KGC} that uses automated test to guide the development of the KG. In order to automate the testing, the SPARQL queries derived from the CQs are wrapped as SHACL~\cite{w3c-shacl} constraints that can be executed and validated. 
Similar to our approach, the "test-driven KG construction" concept uses the feedback from the query execution to uncover incompleteness of the KG or the ontology and to derive improvements. However, there are limitations for functional as well as non-functional requirements that cannot be transformed into SPARQL queries.

\subsubsection{Visual Query Evaluation}
For feedback meetings with domain experts that do not understand SPARQL queries, the answering of CQs can be shown by means of a visualization of the ontology. Relevant classes and relations that are used for answering sub-questions or logical parts of the CQ are highlighted with different colors. The same colors are then used to highlight the corresponding parts, e.g. the triple patterns in the textual SPARQL query. In this way, technical experts can easily explain the relation between a CQ, the derived SPARQL queries and the graph structure of the KG.
Another comprehensible approach are visual SPARQL query builders such as ViziQuer~\cite{viziquer}, a tool for creating and executing visual diagrammatic queries over RDF graphs. In ViziQuer, the user creates visual elements for SPARQL query constructs which are then translated to SPARQL queries and executed.
Query builders for SPARQL can be categorized into form-based query builders, which make use of textual input fields, graph-based query builders, which use visualizations similar to the syntax of the SPARQL query, and natural language-based query builders, which interpret natural language phrases, usually for a limited domain \cite{QueryBuildersUsability}.

\subsubsection{KG Quality Evaluation}
Besides the evaluation of the objectives, the KG instance is evaluated in terms of quality. The evaluation dimensions of KG quality can be summarized to accuracy, completeness, consistency, timeliness, trustworthiness, and availability \cite{KG-quality}. Wang et al.~\cite{KG-quality} discuss these quality dimensions for KGs and give an overview over measures for their improvement.
Based on the results of the quality control, measures should be taken immediately or during the next iteration of the KG construction since the procedure model does not schedule a specific step for the quality enhancement of a constructed KG. Following the agile approach, quality control is intrinsic to each step of each iteration of the KG construction.
We propose a practical quality check of the RDF KG on three levels: The syntax and grammar of the RDF graphs generated from RDF mapping are validated by parsing the Turtle or N-Quads output files. The syntax and grammar are usually validated when importing the RDF files into a Triple Store. The second level is a search for logical inconsistencies in the RDF graph by means of an OWL reasoner. The complexity of the logical verification depends on the OWL profile that has been chosen for ontology modeling. We recommend the OWL-LD rule set \cite{owl-ld}, a subset of the OWL RL profile in the Linked Data community, for identifying logical inconsistencies. On the third level, the structure and content of the RDF graph are evaluated using RDF shapes languages such as SHACL~\cite{w3c-shacl} or ShEx~\cite{shex}. As described in Sec.~\ref{section_proceduremodel_modeling}, RDF shapes can be used in addition to the RDFS/OWL ontology to define shape constraints on a schema level. KG cleaning and error detection are achieved by verifying domain-specific patterns expressed in graph shapes \cite{KGlifecycle}.

\subsection{Deployment} \label{section_proceduremodel_deployment}
The final step "deployment" has a position outside of the circular process in figure~\ref{fig_proceduremodel} to illustrate its special role. Different to the previous steps, the process for deploying the KG is not mandatory in each iteration of the procedure model. Moreover, the operation of a KG is a permanent process which runs parallel to the steps for extending or enhancing the KG. The characteristics of the step "deployment" are listed in table~\ref{table_step_deployment}. The process step summarizes every task that is necessary to deploy, operate and maintain the KG for the industry application. This process step is intentionally called "deployment" to distinguish it from the publication of openly available KGs or Linked Open Data on the Web.

\begin{table}[h!]
\begin{threeparttable}
\label{table_step_deployment}
\caption{Characteristics of the procedure step "deployment"}
\centering
\setlength{\tabcolsep}{2pt}
\setlength{\parskip}{0pt}
\begin{tabular}{ m{0.23\textwidth} m{0.74\textwidth} }
\hline
Goals\newline &
\begin{itemize}
    \item production version of the KG
    \item integration with applications and existing systems
    \item operation and maintenance
\end{itemize}
\\[-8pt]
\hline
Work methods\newline &
\begin{itemize} 
    \item finalize the KG so that it can be used by stakeholders
    \begin{itemize}[nosep,leftmargin=10pt,label={--}]
         \item deployment into existing infrastructure
         \item setup interface and security for applications
         \item (performance) testing
    \end{itemize}
    \item automation of the construction process: data extraction, processing, transformation, import and validation
\end{itemize}
\\[-8pt]
\hline
Stakeholders\newline &
\begin{itemize} 
    \item technical experts for graph database
    \item IT experts for the infrastructure
    \item (optional) app developers
\end{itemize}
\\[-8pt]
\hline
\vspace{3pt}Results\newline &
\begin{itemize} 
    \item operational knowledge graph
\end{itemize}
\\[-8pt]
\hline
\vspace{3pt}Tool support\newline &
\begin{itemize} 
    \item Continuous Integration (CI) and Continuous Delivery (CD) tools
    \item Cloud services, e.g. SaaS
\end{itemize}
\\[-8pt]
\hline
\vspace{3pt}Difficulties\newline &
\begin{itemize} 
    \item good performance in case of high data volume or mass requests
    \item maintenance, update and extension during operation
\end{itemize}
\\[-8pt]
\hline
General advice\newline &
\begin{itemize} 
    \item requirements for "Big Data" applications should be gathered in "business understanding", considered during "modeling" and "graph setup"
    \item test and implement a life cycle strategy before deployment
\end{itemize}
\\[-8pt]
\hline
\end{tabular}
\end{threeparttable}
\end{table}

\vspace*{-1.0\baselineskip}

\subsubsection{KG for Production}
There are major differences between the proof-of-concept KG instance from section~\ref{section_proceduremodel_graphsetup} and the KG for production. 
In the step "graph setup", the KG instance is built and tested manually. The production version uses automated processes, continuous integration (CI) and/or continuous delivery (CD) for data processing, graph setup and validation. Governance and versioning methods are required for generating consistent and reproducible results from e.g. ETL scripts, ontology documents, RDF mapping rules and SPARQL queries. For example, the artefacts from the ontology development process can be used to generate Application Programming Interfaces (APIs) for application developers to simplify the consumption of KG data \cite{onto-api}.
Furthermore, applications and users demand better performance from the production version of the KG, e.g. low query execution times. Depending on the industry use case, the deployment has specific requirements for performance that which may require Cloud solutions and distributed architectures that scale to a high volume of RDF data \cite{rdf-in-the-cloud}. 

\subsection{Lean and Agile Development}

One of the biggest advantages of KGs is the flexibility of the graph structure. In contrast to relational data models, the graph structure can be changed or extended without great expense. In RDF KGs, new classes or properties are simply added to the graph because data and schema are modeled on the same level. For this reason, KGs are inherently well suited for agile development.

The KG-PM from Fig.~\ref{fig_proceduremodel} incorporates lean principles which regard everything that is not adding value to the customer as waste. In the context of KG creation, the customers are the later users of the (enterprise) KG. The users' business goals are defined by means of CQs in the first step of the procedure model. These CQs are used throughout the process to control the scope and quality of the KG development. In order to reduce unnecessary work, each following work step is limited to the scope of the prioritized CQs. Only relevant data sources are selected and only necessary input data is prepared in the steps "data understanding" and "data preparation". During "modeling", a minimal data model is designed and implemented in a minimal ontology. The ontology is limited to classes, properties, individuals and axioms required for answering the selected CQs. By implementing only a small increment of the KG in each iteration, the team is able to deliver valuable extensions of the KG more early.

The efficient creation of working increments is a key component of agile software development. There are many known advantages to breaking down product development into small work packages which require less amount of up-front planning and design. Most advantages over sequential processes like the waterfall model can be derived from the short feedback loop and adaptation cycle. In the KG-PM, the step "evaluation" (Sec.~\ref{section_proceduremodel_evaluation}) is used to evaluate the answerability of the CQs and for obtaining feedback from domain experts and potential users. After the evaluation phase, a new iteration of the procedure starts. In the new iteration's first step, the collection of CQs is updated based on the insight from the evaluation or from new requirements. In accordance with the values of the manifesto for agile software development \cite{agile_manifesto}, the process is designed to favor responding to change over following a plan.
The presented KG-PM can be aligned with Scrum \cite{scrum}, the most popular agile framework in software development. The KG development is broken down into work packages with a specific goal and implemented within time-boxed iterations over the circular process shown in Fig.~\ref{fig_proceduremodel}. Each iteration represents a sprint with a duration of one to four weeks. The collection of CQs constitutes the product backlog which is managed by the Scrum product owner. During sprint planning, the Scrum team selects the sprint goal and the CQs that should be answered by the KG at the end of the upcoming development iteration. As described in section~\ref{section_proceduremodel_evaluation}, the goals are evaluated by checking if the derived SPARQL queries answer the selected CQs. Depending on the results of the review, the CQ backlog is extended, modified or prioritized in a different way. In this way, the KG-PM is aligned with the three pillars of Scrum -- transparency, inspection and adaptation.

\section{Related Work} \label{section_relatedwork}
The construction of a KG involves information systems, software components, data, and human intelligence. Thus, many different methodologies for organizing the KG construction (KGC) and ontology development have been proposed.
Approaches for ontology development include the Cyc, DILIGENT, Grüninger and Fox methodology, the KACTUS approach, METHONTOLOGY \cite{onto-methods-survey} or Ontology Development 101 \cite{ontology101}, which is used in the step "modeling" of the presented KG-PM. While ontologies primarily focus on creating knowledge models, KG comprise large amounts of data connected via knowledge. Within KG-PM, ontology design is one specific step, described in Sec.~\ref{section_proceduremodel_modeling}.

\subsubsection{Consolidated KG Development Process}
\begin{figure}[ht]
    \setlength{\abovecaptionskip}{6pt}
    \setlength{\belowcaptionskip}{-10pt}
    \centering
    \includegraphics[width=1.0\textwidth]{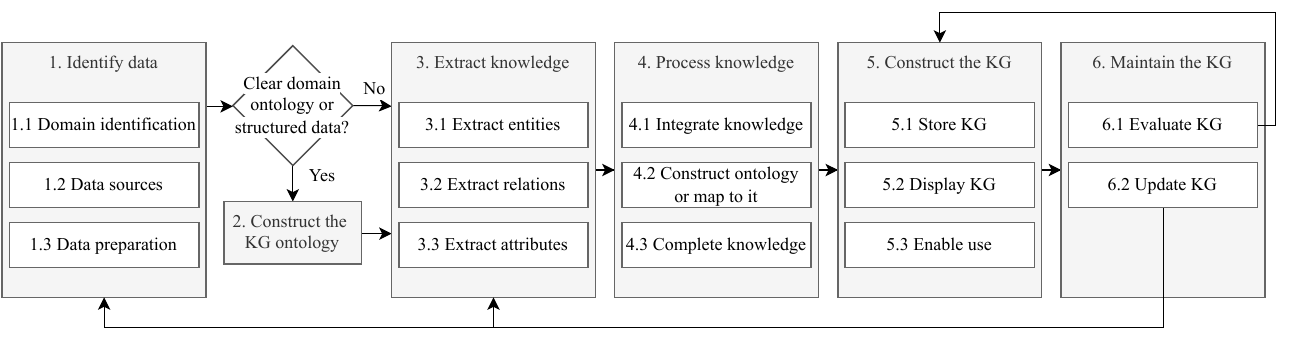}
    \caption{Consolidated KG development process by \cite{KG-devprocess-survey}; replica figure with sub-steps added to step "identify data"} 
    \label{fig_kgdevsynthesis}
\end{figure}

Tamašauskaite and Groth \cite{KG-devprocess-survey} performed a synthesis of common steps in KG development based on the results of a systematic literature analysis. The derived process for KG construction consists of the following six steps that are illustrated in figure~\ref{fig_kgdevsynthesis}: Identify data, construct the KG ontology, extract knowledge, process knowledge, construct the KG, and maintain the KG.
The process incorporates both top-down and bottom-up approaches in terms of ontology development. In the top-down approach, the ontology is defined up-front before knowledge extraction. The great majority of processes use the bottom-up approach in which knowledge is extracted from data before modeling the ontology.

The KG-PM presented in this paper follows the bottom-up procedure. The domain ontology is developed based on the expert knowledge and the results of the data examination at the beginning of the KG-PM. In comparison with our process from section~\ref{section_proceduremodel}, the consolidated process by \cite{KG-devprocess-survey} is focused on extracting and discovering unknown knowledge from the data used for KG construction while our procedure model has the goal of integrating information for a specific application.
The consolidated process starts with data identification which contains tasks for defining the domain, finding data sources, choosing the appropriate data acquisition method and preparing the data. Three separate steps are dedicated to "business understanding", "data understanding" and "data preparation" in the KG-PM. We place more emphasis on the definition of the use case and the preparation of data to cope with the complexity of industry applications and industrial data. For example, machine data requires some kind of translation or analysis to extract knowledge, e.g. RDF statements about the operating state of the machine.
In the third step of the consolidated process, knowledge is derived from the data using techniques such as named entity recognition or OIE methods. These techniques extract RDF statements from text documents, typically loaded from the Web. In our experience with industry applications, the majority of data sources is structured data in which entities, relations and attributes are defined -- even though they are are missing semantics. Instead of OIE, valuable information is extracted from structured data and knowledge shared by the domain experts.
The following process steps for KG construction, querying, evaluation and update are structured differently than the steps of our KG-PM, but contain very similar tasks.

\subsubsection{KG Lifecycle Process Model}
Fensel et al. \cite{FenselKG_HowToBuild} define an overall process model for building a KG that consists of four major steps: knowledge creation (1), knowledge hosting (2), knowledge curation (3), and knowledge deployment (4). These four process steps represent the life cycle of a KG \cite{KGlifecycle}.
The process begins with a knowledge acquisition phase, which is also called knowledge engineering. The presented methodology for semantic annotation uses a combination of a bottom-up and a top-down approach for domain-specific modeling of knowledge. Manual and semi-automatic editing, RDF mapping techniques and automatic annotations using natural language processing (NLP) and machine learning (ML) are used to create RDF data for the KG. In the second step, the semantically enriched data is stored in a graph-based repository, typically a graph database. The following process for "assessing and improving a knowledge graph in various dimensions, especially correctness and completeness" is named knowledge curation \cite{KGlifecycle}. The curation step has the sub-steps knowledge assessment, cleaning and enrichment. During the KG assessment, the quality and usefulness of the KG is rated. The goal of "knowledge cleaning" is to improve the correctness of the KG by detecting and fixing errors, e.g. by verifying domain-specific patterns expressed as SHACL shapes \cite{KGlifecycle}. The sub-step "knowledge enrichment" aims at improving the completeness of the KG by adding new statements. Techniques such as record linkage are used to resolve different identifiers that refer to the same entity. The last process step describes the deployment and use of the KG.
In comparison to the KG-PM presented in this paper, the four-step process model by \cite{FenselKG_HowToBuild,KGlifecycle} is a general solution that can be applied to any domain.

\subsubsection{Process Models for KGC in Academia and Industry}
Ryen et al. \cite{KGDevPubSurvey} reviewed publications from the Extended Semantic Web Conference, the International Semantic Web Conference, the Journal of Web Semantics, and the Semantic Web Journal about the creation of a KG from structured and semi-structured data using Semantic Web technologies. The majority of the reviewed studies on KG originate from the public domain, presumably because the industry is reluctant to make their solutions public. 
Ryen et al. \cite{KGDevPubSurvey} summarize the tasks for KGC into five unordered phases: ontology development, data preprocessing, data integration, quality and refinement, and data publication. The study found that ontology reuse, URI/IRI strategy, data linking, RDF transformation,
and publication are the most common tasks while tasks for evaluation, validation or versioning are mentioned less often.

Companies who provide solutions for knowledge graphs often present practical process models for building KGs.
The US company Stardog describes the following iterative process for KG construction that has many similarities with the presented KG-PM \cite{stardog_HowtoKG}:
\begin{itemize}[leftmargin=24pt,topsep=2pt]
  \item determine the use cases,
  \item outline the necessary data,
  \item organize the data,
  \item map relationships with ontologies,
  \item create KG from a small data sampling,
  \item analyze the deliverable and adjust,
  \item expand the KG (turn into EKG),
  \item evaluate the business needs.
\end{itemize}
The process described in Stardog's article \cite{stardog_HowtoKG} has a similar emphasis on the delivery of a working KG for specific use cases. The lean and agile construction of the KG should be tailored to the selected use cases and avoid unnecessary modeling. In both procedure models, subject matter experts help to define business questions and identify relevant data sources. After building the first KG "deliverable", its usefulness and accuracy is analysed with respect to the business questions and use case.
Ontotext, the company behind the GraphDB database, presents ten steps for building a KG \cite{ontotext_HowtoKG}. The first four steps for defining business questions (1), gathering relevant data (2), data cleaning (3), and creating the semantic model (4) match with our KG-PM. Instead of the process step "graph setup" from Sec.~\ref{section_proceduremodel_graphsetup}, Ontotext's approach has four steps for RDF mapping or virtualization (5), data reconciliation, fusion and alignment (6), the architecture of the data management and search layer (7), and KG enrichment via reasoning and analysis (8). The KG-PM omits an explicit step for KG enrichment, as this is part of a new iteration of the procedure model. After that, the usability is evaluated (9) by finding answers to the original business questions. Similar to the process by \cite{KG-devprocess-survey}, the final step is dedicated to the maintenance and updates of the KG (10).

\vfill

\section{Conclusion}
This paper presents a procedure model for building (enterprise) knowledge graphs (KG) based on W3C Semantic Web standards. Requirements and experiences from industry application have been built into the seven steps of the procedure model: business understanding, data understanding, data preparation, modeling, graph setup, evaluation, and deployment. The characteristics of each process step are defined and related to existing work or design patterns.

In future work, these process steps can be expanded to include automation options. The progress in the area of generative AI and Large Language Models creates new methods for automating tasks in the different steps of the KG construction, from information extraction to query building. Even though AI tools are unable to understand complex industry processes on their own, they serve as a powerful tool for subject matter experts during the development as well as the usage of the KG.

\paragraph{{\normalfont\textbf{Acknowledgments.}}}
Part of this work has been funded by the Bavarian State Ministry of Economic Affairs and Media, Energy and Technology within the research program ``Information and Communication Technology'' (grant DIK0134).

% \bibliographystyle{splncs04}
% \bibliography{literature}

\end{document}